\documentclass[
aps,
prd,
twocolumn,
showpacs,preprintnumbers,nofootinbib,
amsmath,amssymb,
aps,
superscriptaddress,english]{revtex4-1}

\usepackage{amsmath,amssymb,lmodern}
\usepackage{aas_macros}
\usepackage{graphicx}
\usepackage{dcolumn}
\usepackage{bm}
\usepackage{color}
\usepackage{csquotes}
\usepackage[colorlinks=true]{hyperref}
\usepackage{subfigure}
\usepackage{cleveref}
\usepackage{enumitem}
\usepackage{rotating}
\usepackage{multirow}
\usepackage{setspace}
\usepackage{makecell}

\begin{document}

\title{Generic searches for alternative gravitational wave polarizations with 
networks of interferometric detectors}

\author{Peter T.H.~Pang}
\email{thopang@nikhef.nl}
\affiliation{Nikhef -- National Institute for Subatomic Physics, Science Park 105, 1098 XG Amsterdam, The Netherlands}
\affiliation{Department of Physics, Utrecht University, Princetonplein 1, 3584 CC Utrecht, The Netherlands}
\author{Rico K.L.~Lo}
\email{kllo@caltech.edu}
\affiliation{LIGO, California Institute of Technology, Pasadena, California 91125, USA}
\author{Isaac C.F.~Wong}
\email{cfwong@phy.cuhk.edu.hk}
\affiliation{Department of Physics, The Chinese University of Hong Kong, Shatin, N.T., Hong Kong}
\author{Tjonnie G.F.~Li}
\affiliation{Department of Physics, The Chinese University of Hong Kong, Shatin, N.T., Hong Kong}
\author{Chris Van Den Broeck}
\affiliation{Nikhef -- National Institute for Subatomic Physics, Science Park 105, 1098 XG Amsterdam, The Netherlands}
\affiliation{Department of Physics, Utrecht University, Princetonplein 1, 3584 CC Utrecht, The Netherlands}

\begin{abstract}
The detection of gravitational wave signals by Advanced LIGO and Advanced Virgo enables us to 
probe the polarization content of gravitational waves. In general relativity, only tensor
modes are present, while in a variety of alternative theories one can also have vector 
or scalar modes. Recently test were performed which compared Bayesian evidences for the hypotheses
that either purely tensor, purely vector, or purely scalar polarizations were present. Indeed, 
with only three detectors in a network and allowing for \emph{mixtures} of tensor polarizations and 
alternative polarization states, it is not possible to identify precisely which non-standard polarizations 
might be in the signal and by what amounts. However, we demonstrate that one can still infer whether, in 
addition to tensor polarizations, alternative polarizations are present in the first place, 
irrespective of the detailed polarization content. We develop two methods 
to do this for sources with electromagnetic counterparts, both based on the so-called null stream. 
Apart from being able to detect mixtures of tensor and alternative 
polarizations, these have the added advantage that
no waveform models are needed, and signals from any kind of transient source with known
sky position can be used. Both formalisms allow us to combine information from multiple sources so as 
to arrive at increasingly more stringent bounds. For now we apply these on the binary neutron star 
signal GW170817, showing consistency with the tensor-only hypothesis with p-values of 0.315 and 0.790 
for the two methods.
\end{abstract}

\maketitle
\section{Introduction}\label{ns_intro}
Since 2015, Advanced LIGO~\cite{TheLIGOScientific:2014jea} and 
Advanced Virgo~\cite{TheVirgo:2014hva} have been detecting  
gravitational wave (GW) signals on a regular basis
\cite{Abbott:2017xlt,Abbott:2016nmj,TheLIGOScientific:2016pea,Abbott:2017vtc,Abbott:2017gyy,GW170817,Abbott:2017oio,LIGOScientific:2018mvr,Abbott:2020uma}. 
This has enabled a variety of tests of general relativity (GR), including but not limited to
the strong-field dynamics of binary coalescence 
\cite{TheLIGOScientific:2016src,Abbott2016e,LIGOScientific:2019fpa,Abbott:2018lct}, 
the way GWs propagate over large distances \cite{Abbott_2017_GW_GRB,Abbott:2017vtc,LIGOScientific:2019fpa}, 
and preliminary investigations into their polarization content \cite{Isi:2017fbj,Abbott:2017oio,Abbott:2018lct}.  

Generic metric theories of gravity allow for the existence of up to six polarization states 
for gravitational waves \cite{PhysRevD.8.3308}, which can be categorized into tensor modes, 
vector modes, and scalar modes. While GR only permits the tensor modes, some theories of gravity predict 
additional polarizations; see e.g.~\cite{PhysRevD.86.022004} and references therein. 
Methodology has been developed to perform searches for alternative polarizations in continuous 
gravitational wave signals \cite{Isi:2015cva,Isi:2017equ,Abbott:2017tlp} as well as stochastic
backgrounds \cite{Nishizawa:2009bf,Nishizawa:2009jh,Nishizawa:2013eqa,Callister:2017ocg,Abbott:2018utx}. 

In the case of signals from coalescing compact binaries, in 
\cite{TheLIGOScientific:2016src,Isi:2017fbj,Abbott:2017oio,Abbott:2018lct}, 
ratios of Bayesian evidences were 
computed for the hypotheses that
only tensor polarizations, only vector polarizations, or only scalar polarizations were present in 
the signals. Yet, in realistic alternative theories of gravity, typically \emph{mixtures} occur 
of tensor modes together with vector and/or scalar polarization states. 

In this paper we 
develop methods that will allow us to check for the existence of such mixtures, in
GW signals from sources whose exact sky position is known through an electromagnetic counterpart. 
As shown by G\"ursel and Tinto \cite{PhysRevD.40.3884}, it is possible to construct a specific 
linear combination of 
the outputs of multiple detectors in a network, the \emph{null stream}, which has the property of
removing any tensor signal that may be present in the data. This idea was further extended and built on in  
\cite{Wen:2005ui,Ajith:2006qk,Chatterji:2006nh}; see 
also \cite{Freise:2008dk,Regimbau:2012ir} in the context of third-generation detectors such 
as Einstein Telescope.   
A commonly used application for LIGO-Virgo gravitational wave searches 
is X-Pipeline \cite{Chatterji:2006nh,xpipeline}, which assumes that only 
tensor polarizations can be present, and then compares the \emph{null energy} (essentially
the square of the null stream) with other 
combinations of detector outputs to search for GW signals that are in accordance with GR.
As pointed out in 
\cite{PhysRevD.86.022004,Hayama:2012au,Hagihara:2018azu,Hagihara:2019ihn,Hagihara:2019rny,Broeck:2013kx}, 
null streams can also be used
to study a signal's non-tensorial polarization content that may result from a GR violation; 
notably, in \cite{Hagihara:2019ihn} an upper bound was put on vector modes 
in GW170817.   

Here we introduce two concrete data analysis pipelines that 
make use of the fact that if there are only tensor polarizations, the 
null energy of \cite{Chatterji:2006nh}, when evaluated at the true sky position, 
follows a particular $\chi^2$ distribution, but 
not if extra polarizations are present. A first method to discover alternative polarization content 
then quantifies to what extent the null energy for the given sky position is consistent with this
$\chi^2$ distribution. In a second method the sky position is \emph{a priori} left free, allowing us to turn 
the tensor-only distribution for the null energy into a probability distribution for
the sky location. This ``sky map" will be biased if alternative polarizations are 
present, which can be quantified by comparing it with the true position of the source on 
the sky. 

Suppose that in a given signal, alternative polarizations 
are in fact present, mixed with tensor polarizations. 
Then to determine the precise nature and relative contributions of the additional modes, 
in general one would need a network of at least five detectors in addition to the sky position
\cite{PhysRevD.86.022004,Hayama:2012au,Broeck:2013kx,Takeda:2018uai}.\footnote{An exception occurs for certain 
special sky positions with respect to the network; 
see \cite{Hagihara:2018azu,Hagihara:2019ihn,Hagihara:2019rny}. In the case of 
third-generation detectors such as Einstein Telescope and Cosmic Explorer, 
where signals from coalescing binaries will be in band for an extended period of time, 
the variation in time of the antenna patterns can also be used \cite{Takeda:2019gwk}.}
Although in the near future KAGRA \cite{Aso:2013eba} will join the discovery efforts, and 
LIGO-India \cite{M1100296} is about to be built, 
for now only the two LIGO interferometers and Virgo are making regular detections. 
However, what we want to establish first of all is whether or not GW signals contain non-standard 
polarizations, \emph{irrespective of how much each possible type contributes}, and this is what our two 
methods enable us to do. If we were to find evidence that GW signals tend to contain alternative 
polarizations, then this would be a powerful motivation to extend the global detector network even 
further, in order to be able to study what precisely is contained in a mixture of polarizations. 
Note that this should include checking whether tensor modes are in fact in the mix. 

Finally, the fact that our methodology is based on the null energy implies that no waveform models 
are required, so that apart from compact binary coalescences, signals from any 
transient source (supernovae, cosmic strings, ...) can be studied, on condition that 
the sky position is known, e.g.~through an identifiable electromagnetic counterpart. 

This paper is structured as follows. Sec.~\ref{sec:polarization} recalls  
the effects of different polarization modes on  interferometric 
gravitational wave detectors. Sec.~\ref{sec:formulation} 
explains our two methods for finding additional polarizations, one based on the null energy for the 
true sky position, and the other on sky maps. In Sec.~\ref{sec:result} we perform a simulation whereby 
signals with a varying amount of 
scalar polarization in addition to the tensor modes are ``injected" into synthetic stationary, Gaussian
noise, and we compare the performance of the two analysis pipelines. The methodology
is also applied to the binary neutron star signal GW170817, showing consistency with the hypothesis that
only tensor polarizations were present. A summary and conclusions are given in Sec.~\ref{sec:conclusion}.

\section{Gravitational wave polarizations}
\label{sec:polarization}
\begin{figure}[!h]
	\begin{center}
		\includegraphics[width=0.8\linewidth]{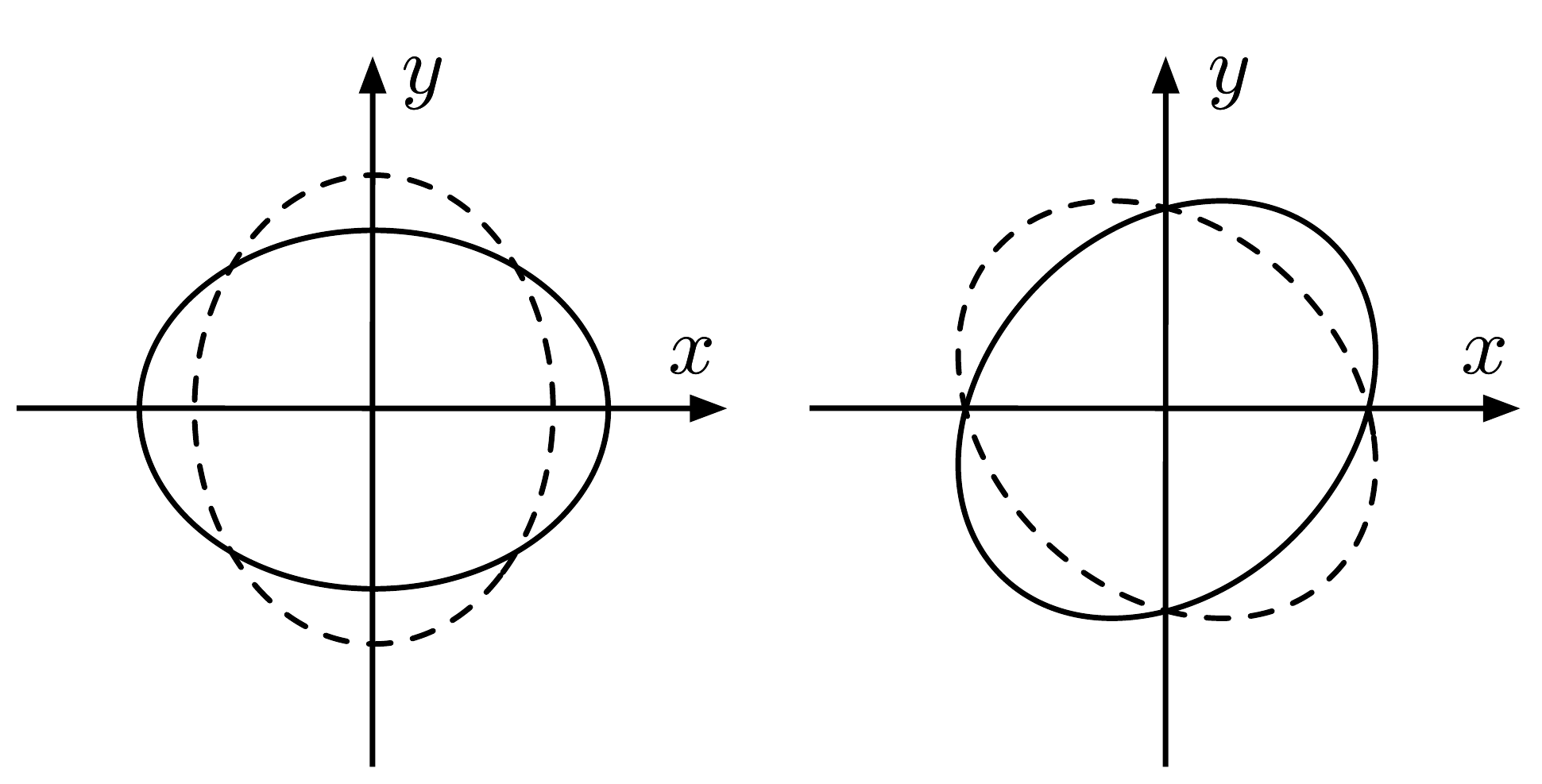}
		\includegraphics[width=0.8\linewidth]{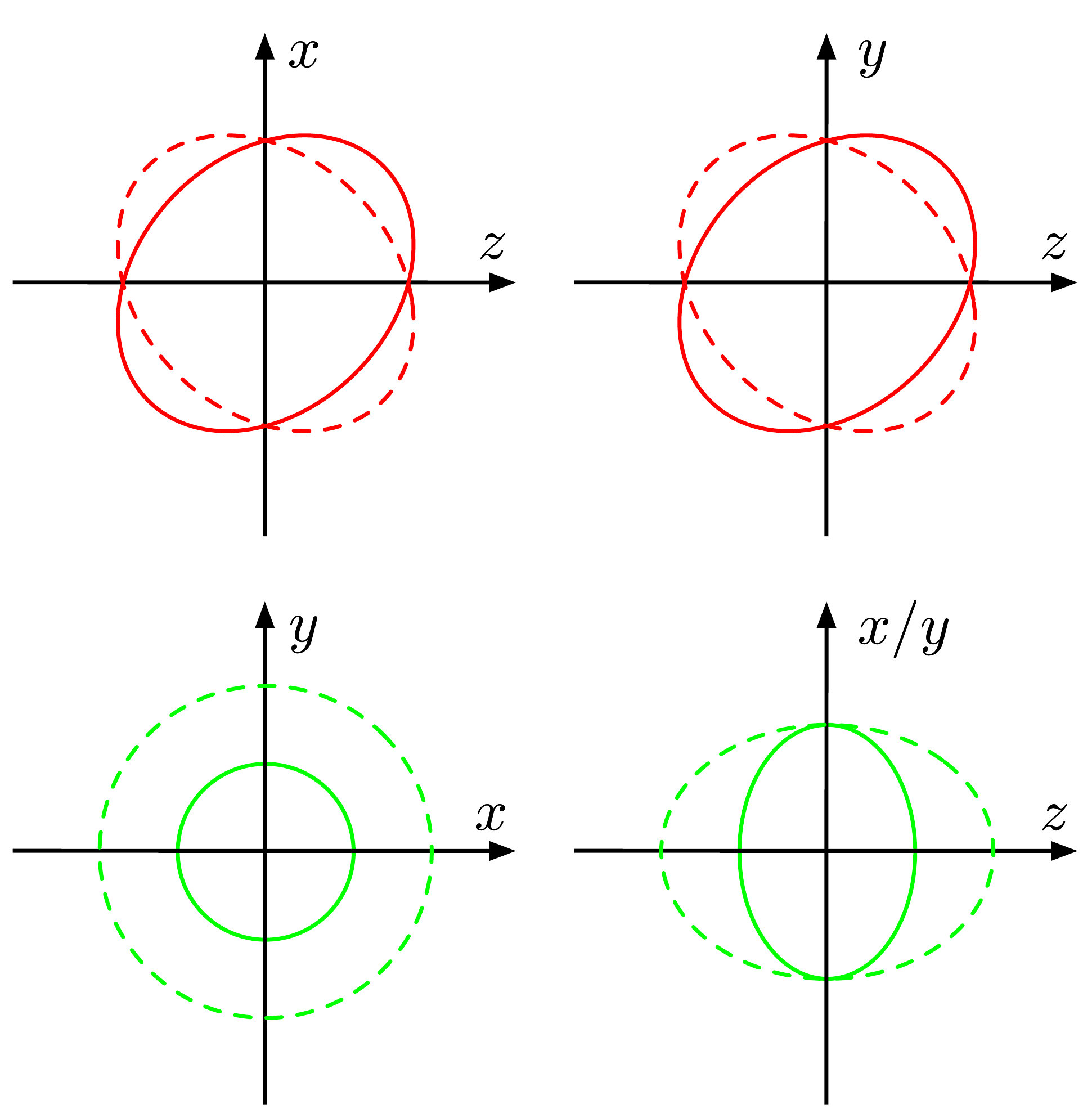}
	\end{center}
	\caption[Effects of gravitational waves with non-general-relativity polarization modes on a 
	ring of test mass]{The effect on a ring of free-falling test particles of a gravitational 
	wave in ``+" tensor mode (upper left), ``$\times$" tensor mode (upper right), 
	``X" vector mode (middle left), ``Y" vector mode (middle right), breathing mode (lower left) 
	and longitudinal mode (lower right). 
	In each case the wave is traveling in the $z$-direction. The solid and dotted lines 
	are the states of the ring with a phase difference of $\pi$.}
	\label{fig:extra_polar_demo}
\end{figure}

In generic metric theories of gravity, up to six independent polarization modes can be present, 
namely a breathing mode, a longitudinal mode, the ``X" vector mode, the ``Y" vector mode, and the 
usual tensor modes predicted by GR \cite{sse}. The effect of different polarization modes on a ring of 
free-falling test masses is shown in Fig.~\ref{fig:extra_polar_demo}. In all the panels of the figure, a  
gravitational wave is traveling in the $z$-direction. The solid and dotted lines illustrate the deformation
of the ring in response to the various modes. Interferometric gravitational wave detectors will react accordingly, with 
beam pattern functions given by \cite{sse}
\begin{equation}
\begin{aligned}
F_{\text{B}} &= -\frac{1}{2}\sin^2\theta\cos2\phi,\\
F_{\text{L}} &= \frac{1}{2}\sin^2\theta\cos2\phi,\\
F_{\text{X}} &= -\sin\theta(\cos\theta\cos2\phi\cos\psi - \sin2\phi\sin\psi),\\
F_{\text{Y}} &= -\sin\theta(\cos\theta\cos2\phi\sin\psi + \sin2\phi\cos\psi),\\
F_+ &= \frac{1}{2}(1+\cos^2\theta)\cos2\phi\cos2\psi - \cos\theta\sin2\phi\sin2\psi,\\
F_\times &= \frac{1}{2}(1+\cos^2\theta)\cos2\phi\sin2\psi + \cos\theta\sin2\phi\cos2\psi.\\
\end{aligned}
\label{eq:polar_mode}
\end{equation}
Here $(\theta,\phi)$ is the sky location of the source, and $\psi$ is the so-called polarization angle. 
The subscripts ``B'', ``L'', ``X'',``Y'', ``+", and ``$\times$" respectively denote the breathing mode, 
the longitudinal mode, the X vector mode, the Y vector mode, the + tensor polarization, and 
the $\times$ tensor polarization. As can be seen from the expressions for 
$F_{\text{B}}$ and $F_{\text{L}}$, there is a degeneracy 
between the responses of the two scalar modes; in our analyses we only consider the breathing mode.

\section{Methodology}\label{sec:formulation}

Now consider a \emph{network} of $D$ gravitational wave detectors labeled by 
$\alpha = 0, \ldots, D - 1$, located 
on the Earth at positions $\vec{r}_\alpha$ with respect to a geocentric coordinate system, and 
producing strain outputs $d_\alpha$. A gravitational 
wave is assumed to originate from a source with sky location $\hat{\Omega} = (\theta,\phi)$, arriving
at the geocenter at a time $t$. If only the tensor polarizations are present, one has
\begin{equation}
d_\alpha(t+\Delta t_\alpha) 
= F_{+,\alpha}(\hat{\Omega})h_+(t) + F_{\times,\alpha}(\hat{\Omega})h_\times(t) 
+ n_{\alpha}(t+\Delta t_\alpha), 
\end{equation}
where $F_{+, \alpha}$, $F_{\times, \alpha}$ are the beam pattern functions and $n_{\alpha}$ is 
the noise of detector $\alpha$. The time shifts $\Delta t_\alpha$ are given by
\begin{equation}
\Delta t_\alpha = \frac{\vec{r}_\alpha}{c}\cdot(-\hat{\Omega}).
\label{eq:time_shift}
\end{equation}
We can write the $D$-detector observation model more compactly in matrix form:
\begin{equation}
\mathbf{d} = \mathbf{Fh}+\mathbf{n},
\label{eq:d_Fh}
\end{equation}
where
\begin{equation}
\mathbf{d}=
\begin{pmatrix}
d_{0} \\
\vdots \\
d_{D-1}
\end{pmatrix}
\text{,\,\,\,\,\,\,\,}
\mathbf{h}=
\begin{pmatrix}
h_{+} \\
h_{\times}
\end{pmatrix}
\text{,\,\,\,\,\,\,\,}
\mathbf{n}=
\begin{pmatrix}
n_{0} \\
\vdots \\
n_{D-1}
\end{pmatrix},
\end{equation}
and
\begin{equation}
\mathbf{F}=
\begin{pmatrix}
\mathbf{F_{+}} & \mathbf{F_{\times}}
\end{pmatrix}
=\begin{pmatrix}
F_{+,0} & F_{\times,0} \\
\vdots & \vdots \\
F_{+,D-1} & F_{\times,D-1}
\end{pmatrix}.
\end{equation}

\subsection{Null energy}

In the above, the gravitational wave signal $\mathbf{s}=\mathbf{F}\mathbf{h}$ can be viewed as being in a 
subspace of the space of detector outputs spanned by $\mathbf{F_{+}}$ and $\mathbf{F_{\times}}$. 
We can construct the \emph{null projector} $\mathbf{P}_{\text{null}}(\hat{\Omega})$ \cite{xpipeline},  
which projects away the signal when the projector is constructed with the true sky location. The null projector is given by
\begin{equation}
\mathbf{P}_{\text{null}}
=\mathbf{I}-\mathbf{F}_{w}(\mathbf{F}_{w}^{\dagger}\mathbf{F}_{w})^{-1}\mathbf{F}_{w}^{\dagger},
\label{eq:nullprojector}
\end{equation}
where $\dagger$ denotes Hermitian conjugation and $\mathbf{F}_{w}$ are the noise-weighted beam pattern 
functions \cite{xpipeline}. If we apply the null projector with 
the true sky location on the strain data in Eq.~(\ref{eq:d_Fh}), we obtain
\begin{equation}
\begin{aligned}
\mathbf{\tilde{z}}(\hat{\Omega}_{\text{true}})&=\mathbf{P}_{\text{null}}(\hat{\Omega}_{\text{true}})\mathbf{\tilde{d}}_{w}\\
&=\mathbf{P}_{\text{null}}(\hat{\Omega}_{\text{true}})\mathbf{F}_{w}(\hat{\Omega}_{\text{true}})\mathbf{\tilde{h}}
+\mathbf{P}_{\text{null}}(\hat{\Omega}_{\text{true}})\mathbf{\tilde{n}}_{w}\\
&=\mathbf{P}_{\text{null}}(\hat{\Omega}_{\text{true}})\mathbf{\tilde{n}}_{w}
\end{aligned}
\end{equation}
where $\mathbf{\tilde{z}}$ is the \textit{null stream} which only consists of noise living
in a subspace that is orthogonal to the one spanned by $\mathbf{F}_{w, +}$ 
and $\mathbf{F}_{w, \times}$, and $w$ indicates whitening.

In practice, the data are first whitened before applying the null projector. 
As in \cite{xpipeline} we perform the analysis in the time-frequency domain, but 
using the Wilson-Daubechies-Meyer (WDM) 
time-frequency transform because of its superior time-frequency localization \cite{Necula_2012}. 
The \emph{null energy} is then defined as \cite{xpipeline}
\begin{eqnarray}
E_{\text{null}}
&=& \sum_{k}\mathbf{\tilde{z}}_{w}^{\dagger}\mathbf{\tilde{z}}_w \nonumber\\
&=& \sum_{k}\mathbf{\tilde{d}}_{w}^{\dagger}\mathbf{P}_{\text{null}}^{\dagger}\mathbf{P}_{\text{null}}\mathbf{\tilde{d}}_{w} \nonumber\\
&=& \sum_{k}\mathbf{\tilde{d}}_{w}^{\dagger}\mathbf{P}_{\text{null}}\mathbf{\tilde{d}}_{w},
\label{nullenergy_def}
\end{eqnarray}
where $w$ indicates whitening, a tilde refers to the data matrix resulting from the WDM transform, and 
$\sum_k$ sums over the discrete time-frequency pixels. 
The quantity $E_{\text{null}}$ follows 
a $\chi^{2}$ distribution with $\text{DoF} = N_{\tau f}(D-2)$ degrees of freedom, 
where $N_{\tau f}$ is the number of time-frequency pixels used in the analysis.\footnote{Note 
that in \cite{xpipeline}, DoF has an extra prefactor 2, which is not present here because the WDM 
coefficients are real.} 

Now let us assume that there is polarization content in the signal beyond the tensor
polarizations. The whitened data matrix can then be written as
\begin{equation}
\tilde{\textbf{d}}_{w} 
= \textbf{F}_{w,t} {\mathbf{\tilde{h}}}_t + \textbf{F}_{w,e} {\mathbf{\tilde{h}}}_e 
+ \mathbf{\tilde{n}}_{w},
\end{equation}
where the index $t$ is summed over $+$ and $\times$, while the index $e$ is summed over whatever additional
polarizations are present. The null energy calculated at the source's location 
with pure-tensor beam pattern matrix is given by
\begin{eqnarray}
E_{\textrm{null}} &=& \sum_{k} \mathbf{\tilde{d}}_{w}^{\dagger} \mathbf{P}_{\textrm{null}}
\mathbf{\tilde{d}}_{w} \nonumber\\
&=&\sum_{k}\mathbf{\tilde{n}}_{w}^{\dagger}\mathbf{P}_{\textrm{null}} \mathbf{\tilde{n}}_{w} 
	+ \sum_{k}\mathbf{\tilde{h}}_e^{\dagger} \textbf{F}^{\dagger}_{w,e} \mathbf{P}_{\textrm{null}}\textbf{F}_{w,e^{'}}\mathbf{\tilde{h}}_{e^{'}}
\nonumber\\ && + \sum_{k}2\Re(\mathbf{\tilde{h}}_{e}^{\dagger} \textbf{F}^{\dagger}_{w,e} \mathbf{P}_{\textrm{null}}\mathbf{\tilde{n}}_{w}),
\label{Eextra}
\end{eqnarray}
where the last two terms signify the presence of the extra polarizations. Next 
we explain how the $\chi^2$ distribution which the null energy would follow in the absence
of these additional polarizations, can be used to detect them, in two different ways.

\subsection{Null energy method and sky map method}

As mentioned before, we assume gravitational wave events with electromagnetic
counterpart, so that the true sky position $\hat{\Omega}_{\rm true}$ is known. 
With the null energy formalism of the previous subsection, 
this leads to two methods for establishing whether alternative polarizations are present.
\begin{itemize}
\item If there are additional polarizations in the signal, then the null energy evaluated
at $\hat{\Omega}_{\rm true}$ will no longer follow the $\chi^2$ distribution described above. 
To quantify the size of the deviation, we can assign a p-value to the hypothesis that 
only tensor polarizations are present, given by
\begin{equation}
p = \int_{E_{\textrm{null}}}^\infty \chi_{\textrm{DoF}}^2(x) dx,
\label{eq:null_energy}
\end{equation}
where $E_{\rm null}$ is computed from the detector network data $\mathbf{\tilde{d}}_w$ 
and $\hat\Omega_{\rm true}$ assuming the tensor-only 
hypothesis. Under this null hypothesis, $p$ will be distributed uniformly between 0 and 1. 
A small p-value would indicate a strong appearance of the additional terms in the 
right hand side of Eq.~(\ref{Eextra}), suggesting a deviation from GR. In the sequel this method 
will simply be referred to as the \emph{null energy method}. 
\item In the context of null energies, the probability for obtaining particular data 
$\mathbf{\tilde{d}}_w$, given the tensor-only hypothesis $\mathcal{H}_{\rm t}$ and a 
fiducial sky position $\hat\Omega$, can be identified with the probability for the associated 
null energy: 
\begin{equation}
p(\mathbf{\tilde{d}}_w|\hat\Omega, \mathcal{H}_{\rm t}) 
= \chi^2_{\rm DoF}(E_{\rm null}(\mathbf{\tilde{d}}_w, \hat\Omega)),
\label{eq:likelihood}
\end{equation}  
where $\mathcal{H}_{\rm t}$ enters through the construction of $E_{\rm null}$.
Through Bayes' theorem this likelihood function leads to a posterior density for 
the sky position:
\begin{equation}
p(\hat\Omega | \mathbf{\tilde{d}}_w, \mathcal{H}_{\rm t}) 
\propto p(\mathbf{\tilde{d}}_w|\hat\Omega, \mathcal{H}_{\rm t})\,p(\hat\Omega|\mathcal{H}_{\rm t}),
\label{eq:bayes}
\end{equation}
and we let the prior $p(\hat\Omega|\mathcal{H}_{\rm t})$ be uniform on the sphere. 
We can then check for the consistency of the true sky location 
$\hat{\Omega}_{\rm true}$ with the ``sky map" $\mathcal{P}(\hat\Omega) 
\equiv p(\hat\Omega | \mathbf{\tilde{d}}_w, \mathcal{H}_{\rm t})$. 
The point $\hat{\Omega}_{\rm true}$ 
will fall on the boundary of some $(1-q)$ credible 
contour on the sphere, where $q$ is given by 
\begin{equation}
q 
= \int_{\mathcal{P}(\hat{\Omega}) \leq \mathcal{P}(\hat{\Omega}_{\textrm{true}})}
\mathcal{P}(\hat{\Omega})\,d\hat\Omega.
\label{eq:skymap}
\end{equation}
The quantity $q$ is a p-value for the consistency of $\hat\Omega_{\rm true}$ with 
$\mathcal{P}(\hat\Omega)$, which under the tensor-only hypothesis is distributed uniformly between 
0 and 1. In what follows this method will be referred to as the \emph{sky map method}.
\end{itemize}

The two methods are not unrelated. Heuristically, in the null energy method, if $p$ is close to zero then 
from Eq.~(\ref{eq:null_energy}), $E_{\rm null}(\hat\Omega_{\rm true})$ must be large, and 
in the tail of the chi-square distribution. In that case 
$\chi^2_{\rm DoF}(E_{\rm null}(\hat\Omega_{\rm true}))$ will be small, 
and in the sky map method, through Eqs.~(\ref{eq:likelihood})
and (\ref{eq:bayes}), this implies that $\mathcal{P}(\hat\Omega_{\rm true})$ is also small. 
From Eq.~(\ref{eq:skymap}), the value of $q$ will then be small as well. 
Generally speaking, when $E_{\rm null}(\hat\Omega_{\rm true})$
is an outlier with respect to its distribution under the tensor-only hypothesis, we 
can expect $\hat\Omega_{\rm true}$ to be an outlier with respect to $\mathcal{P}(\hat\Omega)$, 
and vice versa. On the other hand, 
the methods do 
differ from each other. In the null stream method, 
the null energy is immediately evaluated at $\hat\Omega = \hat\Omega_{\rm true}$, so that if the 
signal happens to have only tensor modes, the value that $E_{\rm null}$ takes is determined only 
by the noise realization $\mathbf{\tilde{n}}_w$ in the detector network. By contrast, in the sky 
map method we effectively define a likelihood for the full network data $ \mathbf{\tilde{d}}_w$, 
which leads to a distribution for the sky position that is then compared with the true one. However, 
we do not expect major differences in performance: if the signal has strong non-tensorial components, 
both methods will tend to imply an extreme value of $E_{\rm null}(\hat\Omega_{\rm true})$ indicating a 
departure from the hypothesis that only tensor modes are present.

Both methods allow us to combine information from multiple sources so as to 
arrive at a stronger statement on the validity of GR, or lack thereof. 
If GR is an accurate description of the gravitational wave polarization, then 
the values of $p$ obtained from the null energy method and the values of 
$q$ obtained using the sky map method should be distributed uniformly between 
$0$ and $1$. As shown by Fisher \cite{fisher_method}, if one has $N$ samples $\{q_i\}$ 
following a uniform distribution between $0$ and $1$, the test statistic $S$ given by
\begin{equation}
S = -2\sum^N_{i=1}\log(q_i)
\end{equation}
follows a $\chi^2$ distribution with $2N$ degrees of freedom. 
Therefore, the combined p-value $p_{\textrm{com}}$ is given by
\begin{equation}
\label{eq:fisher}
p_{\textrm{com}} = \int_S^\infty \chi^2_{2N}(x)\,dx.
\end{equation}

In what follows, we first test the two methods through simulations,  
and then apply them to the binary neutron star signal GW170817.

\begin{figure*}
	\begin{center}
	\includegraphics[width=\columnwidth]{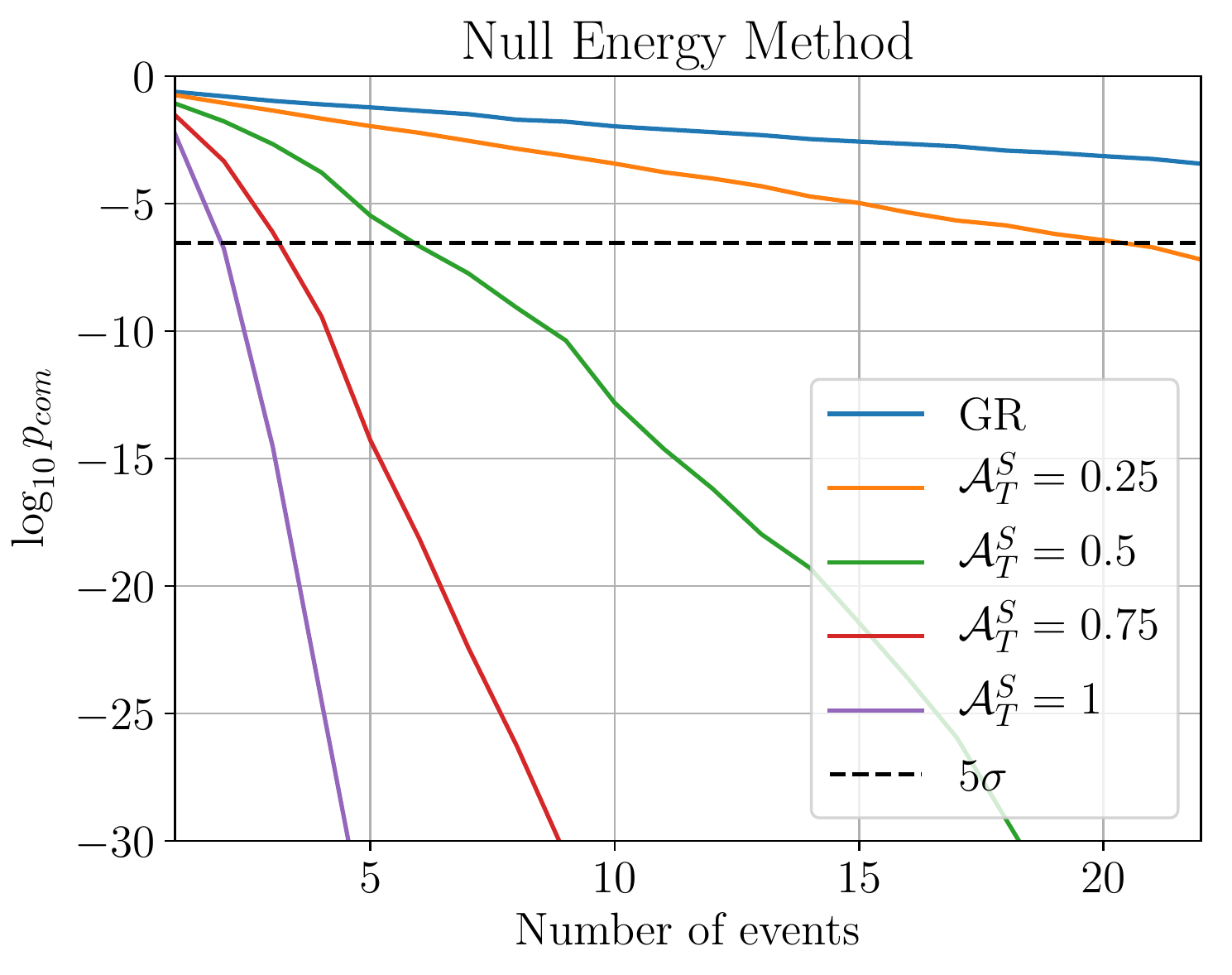} \includegraphics[width=\columnwidth]{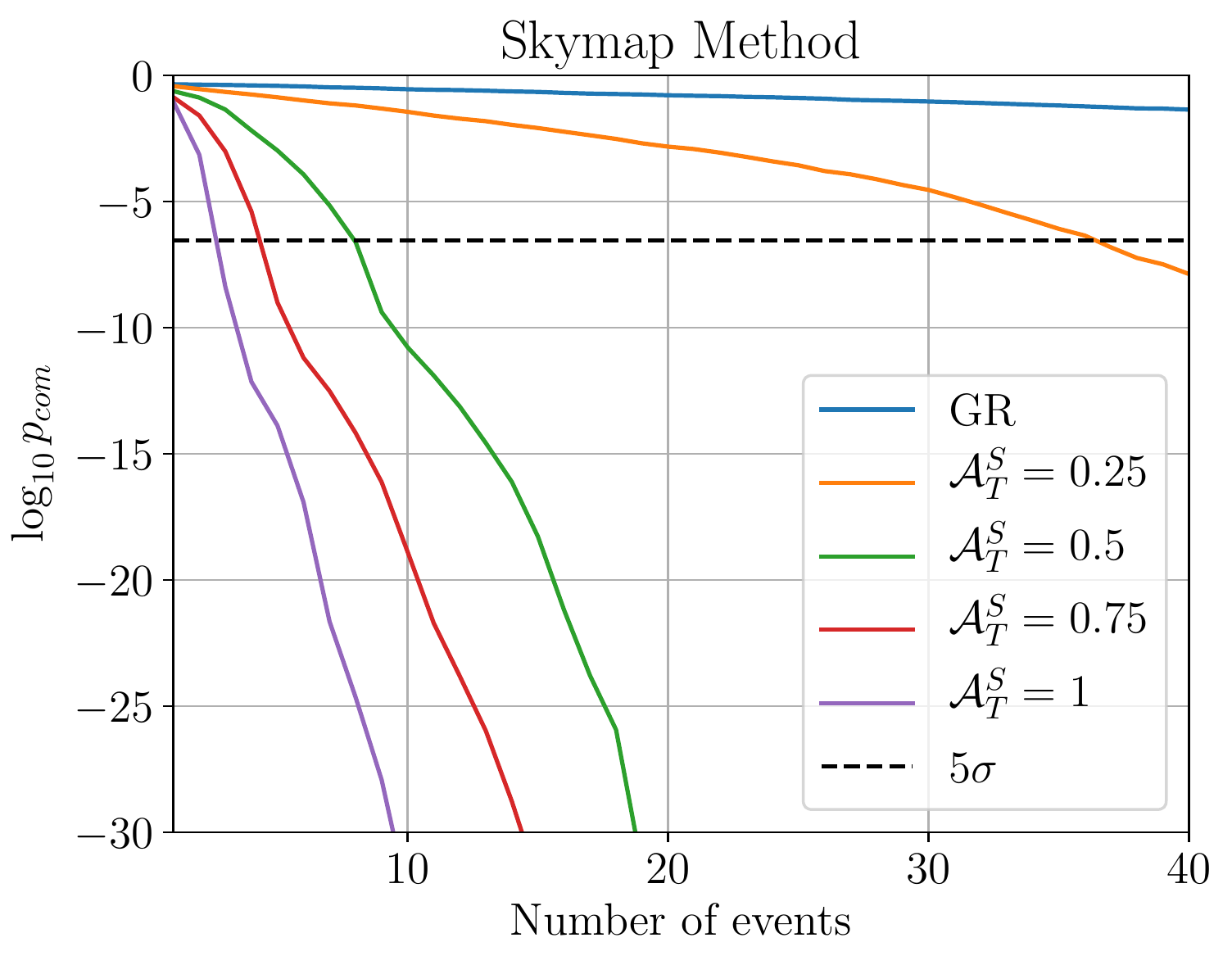}
	\end{center}
	\caption{$\log_{10}$ of combined p-values with null energy method (left panel) and 
	skymap method (right panel) against the number of combined events, 
	both for the GR case and for different sets of mock scalar-tensor signals as described in the main text. 
	For all sets of non-GR injections, GR can be rejected at the $5\sigma$ level 
	with a few to a few tens of detections.}
	\label{fig:pcom_simulated}
\end{figure*}

\section{Simulations, and analyses of GW170817}\label{sec:result}

In this section we evaluate the performance of the null stream and 
sky map methods by ``injecting" simulated signals into synthetic 
stationary, Gaussian noise following predicted noise power spectral densities 
for Advanced LIGO and Advanced Virgo at their respective design sensitivities. 
We take the sources to be zero-spin binary neutron star inspirals with component masses 
uniformly distributed in $[1,2]\,M_\odot$. Positions in the Universe are distributed uniformly in 
co-moving volume up to distances such that the network signal-to-noise remains above 12, and
orientations of the orbital plane are taken to be arbitrary.   
Binary neutron stars are chosen because of their ability to 
generate electromagnetic counterparts when they merge, but in principle
other transient sources could be considered. 

To test the sensitivities of our methods, apart from simulated signals that follow GR 
we also inject sets of mock scalar-tensor waveforms. The scalar component $h_S$ of the latter 
signals is taken to be 
\begin{equation}
h_S(t) = \mathcal{A}^S_T h_T(t; \textrm{with a }\pi/4\textrm{ phase shift}),
\end{equation}
where $h_T$ is the inclination-independent part of the GR polarization $h_+$ (\emph{i.e.}~the part that 
only depends on masses and distance); $\mathcal{A}^S_T$ can be thought of as including
both the inclination dependence of $h_S$ and theory-dependent effects that set the 
intrinsic strength of the scalar component relative to the tensor modes \cite{Takeda:2018uai}. 
The $\pi/4$ phase shift is a strawman for the more 
general ways in which the scalar component's phasing might differ from that of the tensor components; 
in alternative theories of gravity, generically the scalar phase also has a different time 
dependence \cite{PhysRevD.86.022004}. In each of four sets of scalar-tensor injections, 
for simplicity we let $\mathcal{A}^S_T$ take fixed values of 0.25, 0.5, 0.75, or 1.0, effectively taking 
the inclination dependence of $h_S$ to have been averaged over, so that the chosen values for 
$\mathcal{A}^S_T$ can be viewed as indicative of the intrinsic strength of the scalar components relative 
to the tensor modes. 

Fig.~\ref{fig:pcom_simulated} shows $\log_{10}$ of the combined p-value $p_{\textrm{com}}$ 
calculated with the null energy method and the sky map method against the number of events. 
Even for $\mathcal{A}^S_T = 0.25$, it takes only a few tens of detections with electromagnetic
counterparts to establish a $5\sigma$ violation of GR. 
The null energy method appears to be performing slightly better in that it requires fewer sources 
to attain the $5\sigma$ level, although this could be due to the particular parameter values in our 
``catalog" of simulated sources. To make more definitive statements a much larger (and 
computationally costly) simulation campaign would be needed, analyzing many randomly chosen 
``catalogs" of a few tens of sources each. However, on the basis of Fig.~\ref{fig:pcom_simulated} 
we expect that the difference in performance between the methods will not be very pronounced, so 
that the one can be used to complement the other.

We note that $p_{\textrm{com}}$ for GR signals gradually 
approaches the 5$\sigma$ line as the number of events 
increases; this is because of the systematics in the clustering algorithm used. 
The largest high-power cluster is consistently selected to be the event candidate, but 
there is a non-negligible chance for high-power noise pixels to be included in the periphery of the 
cluster. This happens especially when the burst energy of the signal is not significantly higher than the 
background noise. The systematic error accumulates as the number of events increases. 
Hence, when performing analyses for a large number of events, we will have to inject large numbers of
 simulated GR signals into real noise to obtain a reference or ``background" distribution of the 
 test statistic to compare 
``foreground" results with.  This procedure will automatically account for the non-stationary 
and non-Gaussian nature of detector noise as well as the systematics due to the clustering method. 
Nevertheless, the results of Fig.~\ref{fig:pcom_simulated} are already strongly suggestive of 
the sensitivities we can expect to attain, given how rapidly $p_{\rm com}$ becomes small in the 
case of GR violations (note the logarithmic scales on the vertical axes). 

These results show that our analysis pipelines are capable of testing for the existence of alternative
polarization modes in addition to tensor modes with a 3-detector network. 
Given a few tens of detections with known sky positions, both methods are sensitive to a 
scalar component provided that it has appreciable intrinsic strength compared to  
the tensorial components. 

\begin{figure}
	\begin{center}
		\includegraphics[width=1.0\linewidth]{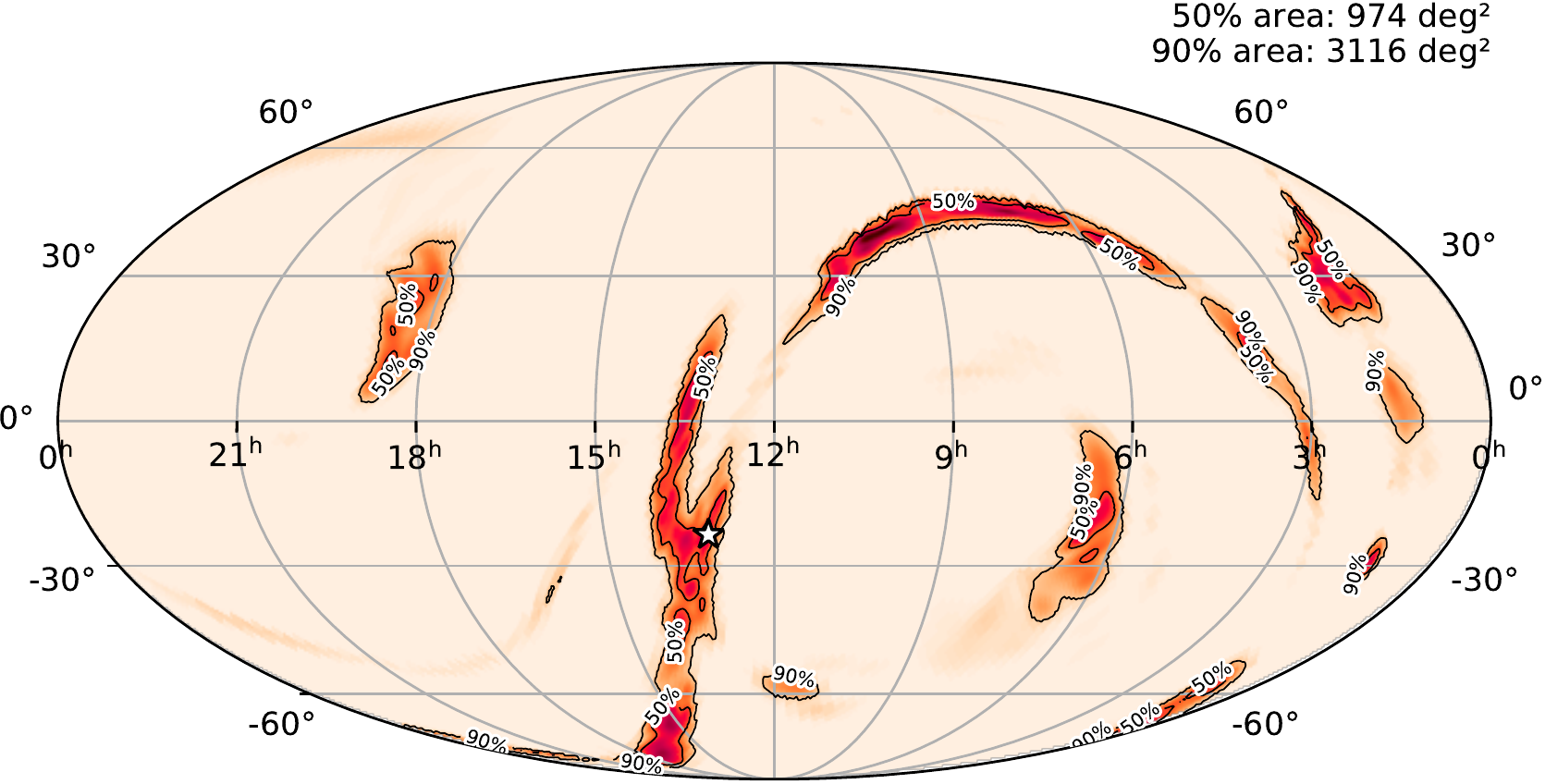}
	\end{center}
	\caption{Sky map $\mathcal{P}(\hat\Omega)$ for GW170817. 
	The star indicates the sky location of the corresponding counterpart 
	SSS17a/AT 2017gfo~\cite{Abbott:2018wiz}. The sky map is 
	consistent with the counterpart location, which is enclosed in the $50\%$
	confidence contour.}
	\label{fig:170817_skymap}
\end{figure}

A binary neutron star coalescence GW170817 was observed on 17 August 2017 
with merger time 12:41:04 UTC (or GPS time 1187008882.4457) \cite{GW170817,GW170817_open_data,Vallisneri_2015}. 
Electromagnetic counterparts were seen, and in particular an optical counterpart was
found with very precise localization at right ascension and declination 
$\alpha=13^{\text{h}}09^{\text{m}}48^{\text{s}}.085\pm0.018$ and $\delta=-23^{\circ}22'53''.343\pm0.218$, 
respectively \cite{Abbott_2017}, which provides us with an opportunity to apply our tests 
for alternative polarizations. The null energy test yields a p-value of 0.315, while 
the sky map method gives 0.790; in the latter case, the sky map $\mathcal{P}(\hat\Omega)$
and true sky location $\hat\Omega_{\rm true}$ are shown in Fig.~\ref{fig:170817_skymap}. 
Hence, neither test allows us to reject the tensor-only hypothesis at anything approaching
the $5\sigma$ level. (We reiterate that properly speaking the p-values should be compared
with a background distribution; this will become important for a larger number of 
events and smaller p-values.)

\section{Summary and conclusions}\label{sec:conclusion}

We have introduced two methods to search for polarization modes 
in addition to the tensor modes, which can be used even with a limited network of 
detectors (e.g.~the two LIGOs and Virgo), though an identifiable electromagnetic counterpart is needed. 
Both formalisms are based on the notion of null energy. In one case (the \emph{null energy test})
we use the statistical distribution of the null energy for the given true sky location
to compute p-values for the validity of the tensor-only hypothesis. The other method
(the \emph{sky map test}) 
first leaves the sky location to be free, turning the distribution of null energy
into a sky map, for which the consistency with the true sky location can again 
be quantified in terms of a p-value. Apart from being able to detect mixtures of
different polarization modes rather than having to consider purely tensor, purely vector, 
and purely scalar hypotheses, no waveform models are needed, so that any transient
gravitational wave signal can be used in the tests, on condition that the sky position of 
the source is known. 

By injecting mock scalar-tensor signals into synthetic stationary and Gaussian noise, 
we illustrated how both tests can find 
scalar contributions at $5\sigma$ confidence with a few tens of signals that have 
electromagnetic counterparts if the scalar contribution is at least at the 25\% level in 
the sense explained above. 
Both methods show a slowly accumulating bias towards a GR violation when applied to 
pure tensor signals, due to the null energy clustering algorithm picking up 
high-energy noise pixels. Thus there is scope for improvement, although even
if tens of signals with counterparts were available today, we would certainly 
be able to already use either method by constructing a reference distribution for our
detection statistic and compare ``foreground" results with this ``background" distribution. 

Finally, we have applied our methods to GW170817, \emph{a priori} allowing for any mixture 
of polarizations, but finding p-values that do not induce us to reject  
the pure-tensor hypothesis.

\section*{Acknowledgments}

We are grateful to the anonymous referee, whose careful reading of the manuscript helped 
us to greatly improve the presentation of the paper.   
PTHP and CVDB are supported by the research program of the Netherlands Organization 
for Scientific Research (NWO).
ICFW and TGFL are partially supported by grants from the Research Grants Council of the 
Hong Kong (Project No. 24304317 and 14306419) and Research Committee of the Chinese University 
of Hong Kong. RKLL and TGFL would also like to 
gratefully acknowledge the support from the Croucher Foundation in Hong Kong.

This research has made use of data, software and/or web tools obtained from the Gravitational 
Wave Open Science Center (https://www.gw-openscience.org), a service of LIGO Laboratory, the 
LIGO Scientific Collaboration and the Virgo Collaboration. LIGO is funded by the U.S. National 
Science Foundation. Virgo is funded by the French Centre National de Recherche Scientifique (CNRS), 
the Italian Istituto Nazionale della Fisica Nucleare (INFN) and the Dutch Nikhef, with contributions 
by Polish and Hungarian institutes.

\bibliography{main}
\end{document}